\documentclass[conference]{IEEEtran}
\usepackage{verbatim}

\usepackage{cite}

\ifCLASSINFOpdf
\else
\fi
\usepackage{color}

\usepackage[cmex10]{amsmath}
\usepackage{amssymb}
\usepackage{amsfonts}
\usepackage{algorithmic}

\usepackage{array}

\usepackage{mdwmath}
\usepackage{mdwtab}

\usepackage{eqparbox}
\usepackage{fixltx2e}

\usepackage{float}

\usepackage{url}

\begin{document}

\title{Good Codes From Generalised Algebraic Geometry Codes}

\author{\IEEEauthorblockN{Mubarak Jibril\IEEEauthorrefmark{1},
Martin Tomlinson\IEEEauthorrefmark{1},
Mohammed Zaki Ahmed\IEEEauthorrefmark{1} and
Cen Tjhai\IEEEauthorrefmark{1}}
\IEEEauthorblockA{\IEEEauthorrefmark{1}School of Computing and Mathematics\\
Faculty of Technology\\ University of Plymouth\\United Kingdom\\ Email: mubarak.jibril@plymouth.ac.uk}}

\maketitle

\begin{abstract}
Algebraic geometry codes or Goppa codes are defined with places of degree one. In constructing generalised algebraic geometry codes places of higher degree are used. In this paper we present $41$ new codes over $\mathbb{F}_{16}$ which improve on the best known codes of the same length and rate. The construction method uses places of small degree with a technique originally published over 10 years ago for the construction of generalised algebraic geometry codes.     
\end{abstract}

\section{Introduction}
In coding theory, it is desirable to obtain an error correcting code
with the maximum possible minimum distance $d$,  given a code length $n$
and code dimension $k$. Algebraic geometry (AG) codes have good
properties and some families of these codes have been shown to be
asymptotically superior as they exceed the well-known Gilbert Vashamov
bound \cite{zink} when the defining finite field  $\mathbb{F}_q$ has
size $q \geq 49$ with $q$ always a square. A closer look at tables of
best known codes in \cite{grassl} and \cite{mint}  shows that algebraic
geometry codes feature as the best known linear codes for an appreciable
range of code lengths for different field sizes $q$. Algebraic geometry
codes are codes derived from curves and were first discovered by Goppa
\cite{goppa} in 1981. Goppa's description uses rational places of the
curve to define these codes. Rational places are called places of degree
one. A generalised construction of algebraic geometry codes was
presented by Xing \textit{et al} in \cite{xing3}\cite{xing4} and Ozbudak
\textit{et al} in \cite{ozbudak}. An extension of the method which utilises places of
higher degrees as well as a concatenation concept was introduced in \cite{xing1}. This method was shown in \cite{san}\cite{ding}\cite{xing2} to be effective in constructing codes that are better than the best known codes and many codes were presented for finite fields up to $\mathbb{F}_9$. In this paper we present several new codes over finite field $\mathbb{F}_{16}$. These codes represent improvements on minimum distance compared to some previously best known codes. We first give a description of the codes and an exposition of the construction in the  Section II.  Finally we present our results in Section III. \\

\section{Construction}
A two dimensional affine space $\mathbb{A}^2(\mathbb{F}_q)$ is given by the set of points $\lbrace (\alpha,\beta):\alpha,\beta\in \mathbb{F}_q \rbrace$
while its projective closure $\mathbb{P}^2(\mathbb{F}_q)$ is given by
the set of equivalence points $\lbrace
\lbrace(\alpha:\beta:1)\rbrace\cup \lbrace(\alpha:1:0)\rbrace\cup
\lbrace (1:0:0) \rbrace : \alpha,\beta\in \mathbb{F}_q \rbrace$. Given a
homogeneous polynomial $F(x,y,z)$, a curve $\mathcal{X}$ defined in
$\mathbb{P}^2(\mathbb{F}_q)$ is a set of distinct points $\lbrace P \in \mathbb{P}^2(\mathbb{F}_q) : F(P)=0\rbrace$. We are only interested in the case where $\mathcal{X}$ is irreducible and is non-singular in order to obtain AG codes. Let $\mathbb{F}_{q^\ell}$ be an extension of the field $\mathbb{F}_q$, the Frobenius automorphism is given as
\begin{align*}
\phi_{q,\ell}&:\mathbb{F}_{{q}^\ell} \rightarrow \mathbb{F}_{q^\ell} &\\
\phi_{q,\ell}&(\beta)=\beta^{q}&\beta \in \mathbb{F}_{q^\ell}  
\end{align*}
and its action on a projective point $(x:y:z)$ in $\mathbb{F}_{q^\ell}$ is 
$$
\phi_{q,\ell}((x:y:z))=(x^q:y^q:z^q).
$$
A place of degree $\ell$ \cite{judy} is a set of $\ell$ points of a
curve defined in the extension field $\mathbb{F}_{q^\ell}$ denoted by
$\lbrace P_0,P_1,\dots,P_{\ell-1}\rbrace $ where each
$P_i=\phi_{q,l}^i(P_0)$. Places of degree one are called rational
places. An example of a place of degree two is a pair of points
$\lbrace P_0,P_1 \rbrace$ such that $P_0=(x,y)$ has coordinates in
$\mathbb{F}_{q^2}$ and $ P_1=\phi_{q,2}(P_0)=(x^q,y^q)$.\\
We will now describe two maps that are useful in the Xing \textit{et al} construction of generalised AG codes. We observe that $\mathbb{F}_q$ is a subfield of $\mathbb{F}_{q^\ell}$ for all $\ell \geq 2$. It is then possible to map $\mathbb{F}_{q^\ell}$ to an $\ell$-dimensional vector space with elements from $\mathbb{F}_q$ using a suitable basis. We define the mapping,
\begin{align*}
\pi_{\ell}&:\mathbb{F}_{{q}^\ell} \rightarrow  \mathbb{F}^\ell_{q} &\\
\pi_{\ell}&(\beta_j)=[c^j_1 \thickspace c^j_2 \dots c^j_l]~~~\beta_j \in \mathbb{F}_{q^\ell}~,~c^j_i \in \mathbb{F}_q.
\end{align*}
Suppose $[\gamma_1 \gamma_2 \dots \gamma_\ell]$ forms a suitable basis of the vector space $\mathbb{F}^\ell_{q}$, then $\beta_j=c^j_1 \gamma_1 + c^j_2 \gamma_2 + \cdots + c^j_{\ell} \gamma_{\ell}$. Finally we use $\sigma_{\ell,n}$ to represent an encoding map from an $\ell$-dimensional message space in $\mathbb{F}_q$ to an $n$-dimensional code space,
\begin{align*}
\sigma_{\ell,n}&:\mathbb{F}^{\ell}_{{q}} \rightarrow  \mathbb{F}^n_{q} &\\
\end{align*}
with $\ell \leq n$.\\
We now give a description of generalised AG codes as presented in
\cite{xing1}\cite{ding}\cite{xing2}. Let $\mathit{F}=F(x,y,z)$ be a
homogeneous polynomial defined in $\mathbb{F}_q$. Let $g$ be the genus
of the curve $\mathcal{X}/\mathbb{F}_q$ corresponding to the polynomial $\mathit{F}$. Also let $P_1,P_2,\dots,P_r$ be $r$ distinct places of
$\mathcal{X}/\mathbb{F}_q$ and $k_i=deg(P_i)$ ($deg$ is degree of).  $W$
is a divisor of the curve $\mathcal{X}/\mathbb{F}_q$ such that $W=P_1+P_2+\cdots+P_r$ and $G$ a divisor so that $supp(W)\cap supp(G)=\varnothing$. More specifically $G=m(Q-R)$ where $deg(Q)=deg(R)+1$. Associated with the divisor $G$ is a Riemann-Roch space $\mathcal{L}(G)$ with $m=deg(G))$ an integer, $m\geq 0$ . From the Riemann-Roch theorem we know that the dimension of $\mathcal{L}(G)$ is given by $l(G)$ and
$$
l(G)\geq m-g+1
$$
with equality when $m\geq 2g-1$. Also associated with each $P_i$ is a $q$-ary code $C_i$ with parameters $[n_i,k_i=deg(P_i),d_i]_{q}$ with the restriction that $d_i \leq k_i$. We denote $\lbrace f_1,f_2,..,f_k:f_l \in \mathcal{L}(G)\rbrace$ as a set of $k$ linearly independent elements of $\mathcal{L}(G)$ that form a basis. We can create a generator matrix for a generalised AG code as such,

$$
M=
\left [ \begin{smallmatrix}
\sigma_{k_1,n_1}(\pi_{k_1}(f_1(P_1))) & \dots\dots &\sigma_{k_r,n_r}(\pi_{k_r}(f_1(P_r))) \\
\sigma_{k_1,n_1}(\pi_{k_1}(f_2(P_1))) &\dots \dots &  \sigma_{k_r,n_r}(\pi_{k_r}(f_2(P_r)))\\
            .                         & . & .\\
            .                         & . & .\\
            .                         & . & .\\
\sigma_{k_1,n_1}(\pi_{k_1}(f_k(P_1))) & \dots \dots &  \sigma_{k_r,n_r}(\pi_{k_r}(f_k(P_r)))
\end{smallmatrix} \right]
$$
 where $f_l(P_i)$ is an evaluation of a polynomial and basis element
 $f_l$ at a point $P_i$, $\pi_{k_i}$ is a mapping from
 $\mathbb{F}_{q^{k_i}}$ to $\mathbb{F}_q$ and $\sigma_{k_i,n_i}$ is the
 encoding of a message vector in $\mathbb{F}^{k_i}_{q}$ to a code vector
 in $\mathbb{F}^{n_i}_{q}$. It is desirable to choose the maximum
 possible minimum distance for all codes $C_i$ so that $d_i=k_i$. The
 same code is used in the map $\sigma_{k_i,n_i}$ for all points of the
 same degree $k_i$ i.e. the code $C_j$ has parameters $[n_j,j,d_j]_q$
 for a place of degree $j$. Let $A_j$ be an integer denoting the number
 of places of degree $j$ and  $B_j$ be an integer such that $0\leq B_j
 \leq A_j$. If $t$ is the maximum degree of any place $P_i$ we choose to
 use in the construction, then the generalised AG code is represented as
 a $C(k;t;B_1,B_2,\dots,B_t;d_1,d_2,\dots,d_t)$. Let $[n,k,d]_q$
 represent a linear code in $\mathbb{F}_q$ with length $n$, dimension
 $k$ and minimum distance $d$, then a generalised AG code is given by
 the parameters \cite{xing1},
\begin{align*}
k&=l(G)\geq m-g+1\\
n&=\sum_{i=1}^{r}  n_i=\sum_{j=1}^t B_jn_j \\
d&\geq\sum_{i=1}^{r} d_i-g-k+1=\sum_{j=1}^t B_jd_j-g-k+1.
\end{align*}

\section{Results}
 We use two polynomials and their associated curves to obtain codes in
 $\mathbb{F}_{16}$ better than the best known codes in \cite{mint}. The
 two polynomials are given in  Table  \ref{poly2} while Table \ref{poly}
 gives a summary of the properties of their associated  curves (with $t=4$). The number of places of degree $j$, $A_j$, is determined by computer algebra system {\small MAGMA}~\cite{magma}. The best known linear codes from \cite{mint} over $\mathbb{F}_{16}$ with $j=d_j$ for $1 \leq j\leq 4$ are
\begin{equation*}
[1,1,1]_{16}~~~[3,2,2]_{16}~~~[5,3,3]_{16}~~~[7,4,4]_{16}
\end{equation*}
which correspond to $C_1$, $C_2$, $C_3$ and $C_4$ respectively. Since $t=4$ for all the codes in this paper and 
$$
[d_1,d_2,d_3,d_4]=[1,2,3,4]
$$
we shorten the representation \\
$$
C(k;t;B_1,B_2,\dots,B_t;d_1,d_2,\dots,d_t) \equiv C(k;B_1,B_2,\dots,B_t).
$$
Tables \ref{f1}-\ref{f2} give codes obtained from the two curves
associated with the two polynomials $F_i$ for $1\leq i\leq 2$ that
improve on the best constructible codes in the tables in
\cite{mint}. Table \ref{f3} gives new codes that improve on both
constructible and non-constructible codes in \cite{mint}. It is also worth noting that codes of the form $C(k;N,0,0,0)$ are simply Goppa codes (defined with only rational points). The symbol \# in the Tables \ref{f1}-\ref{f2} denotes the number of new codes from  each generalised AG code $C(k;B_1,B_2,\dots,B_t)$ . The tables in \cite{geer2} contain curves known to have the most number of rational points for a given genus. Over $\mathbb{F}_{16}$ the curve with the highest number of points with genus $g=12$ from \cite{geer2} has $88$ rational points, was constructed using class field theory and is not defined by an explicit polynomial. On the other hand the curve $\mathcal{X}_{1}/\mathbb{F}_{16}$ obtained by Kummer covering of the projective line in \cite{shabat} has $A_1=83$ rational points and genus $g=12$ and is explicitly presented. Codes from this curve represent the best constructive codes in $\mathbb{F}_{16}$ with code length $83$. The curve $\mathcal{X}_{2}/\mathbb{F}_{16}$ is defined by the well-known Hermitian polynomial. 
\begin{table}[!t]
\caption{Polynomials in $\mathbb{F}_{16}$}
\label{poly2}
\centering

\begin{tabular}{|p{7.5cm}|}
\hline
$\mathit{F}_1=x^5z^{10} + x^3z^{12} + xz^{14} + y^{15}$\\
\hline
$\mathit{F}_2=x^5+y^4z+yz^4$\\
\hline
\end{tabular}
\end{table}

\begin{table}[!t]
\caption{Properties of $\mathcal{X}_i/\mathbb{F}_{16}$}
\label{poly}
\centering
\begin{tabular}{|c|c|c|c|c|c|c|}
\hline
$F(x,y,z)$ & Genus & $A_1$ & $A_2$ & $A_3$ & $A_4$ & Reference\\
\hline
$\mathcal{X}_1$ & $12$ & $83$ & $60$ & $1320$ & $16140$ & \cite{shabat}\\
\hline
$\mathcal{X}_2$ & $6$ & $65$ & $0$ & $1600$ & $15600$ & \\
\hline
\end{tabular}
\end{table}

\begin{table}[!t]
\caption{Best Constructible Codes from $\mathcal{X}_1$}
\label{f1}
\centering
\begin{tabular}{|l|c|l|c|}
\hline
Codes & $k$ Range & Description & \#\\
\hline
$[83,k,d\geq 72-k]_{16}$& $8 \leq k \leq 52$ & $C(k;[83,0,0,0])$& $45$\\
\hline
$[89,k,d\geq 76-k]_{16}$& $9 \leq k \leq 54$ & $C(k;[83,2,0,0])$& $46$\\
\hline
$[94,k,d\geq 79-k]_{16}$& $10 \leq k \leq 57$ & $C(k;[83,2,1,0])$& $48$\\
\hline
$[92,k,d\geq 78-k]_{16}$& $9 \leq k \leq 57$ & $C(k;[83,3,0,0])$ & $49$\\
\hline
$[98,k,d\geq 82-k]_{16}$& $11 \leq k \leq 59$ & $C(k;[83,5,0,0])$ & $49$\\
\hline
\end{tabular}
\end{table}

\begin{table}[!t]
\caption{Best Constructible Codes from $\mathcal{X}_2$}
\label{f2}
\centering
\begin{tabular}{|l|c|l|c|}
\hline
Codes & $k$ Range & Description & \#\\
\hline
$[72,k,d\geq 64-k]_{16}$& $11 \leq k \leq 50$ & $C(k;[65,0,0,1])$ & $40$ \\
\hline
$[79,k,d\geq 68-k]_{16}$& $11 \leq k \leq 48$ & $C(k;[65,0,0,2])$ & $38$ \\
\hline
$[77,k,d\geq 67-k]_{16}$& $10 \leq k \leq 51$ & $C(k;[65,0,1,1])$ & $42$\\
\hline
$[75,k,d\geq 66-k]_{16}$& $9 \leq k \leq 51$ & $C(k;[65,0,2,0])$& $43$\\
\hline
\end{tabular}
\end{table}

\begin{table}[!t]
\caption{New Codes from $\mathcal{X}_1$}
\label{f3}
\centering
\begin{tabular}{|l|c|l|c|}
\hline
Codes & $k$ Range & Description & \#\\
\hline
$[70,k,d\geq 63-k]_{16}$& $10 \leq k \leq 50$ & $C(k;[65,0,1,0])$ & $41$ \\
\hline
\end{tabular}
\end{table}

\begin{table}[!t]
\caption{New Codes in $\mathbb{F}_{16}$}
\label{f4}
\centering
\begin{tabular}{|c|c|c||c|}
\hline
 $N$ & $K$ & $D$ & $D_m$ \\
\hline
$70$ & $10$ & $53$ & $52$\\
\hline
$70$ & $11$ & $52$ & $51$\\
\hline
$70$ & $12$ & $51$ & $50$\\
\hline
$70$ & $13$ & $50$ & $49$\\
\hline
$70$ & $14$ & $49$ & $48$\\
\hline
$70$ & $15$ & $48$ & $47$\\
\hline
$70$ & $16$ & $47$ & $46$\\
\hline
$70$ & $17$ & $46$ & $45$\\
\hline
$70$ & $18$ & $45$ & $44$\\
\hline
$70$ & $19$ & $44$ & $43$\\
\hline
$70$ & $20$ & $43$ & $42$\\
\hline
$70$ & $21$ & $42$ & $41$\\
\hline
$70$ & $22$ & $41$ & $40$\\
\hline
$70$ & $23$ & $40$ & $39$\\
\hline
$70$ & $24$ & $39$ & $38$\\
\hline
$70$ & $25$ & $38$ & $37$\\
\hline
$70$ & $26$ & $37$ & $36$\\
\hline
$70$ & $27$ & $36$ & $35$\\
\hline
$70$ & $28$ & $35$ & $34$\\
\hline
$70$ & $29$ & $34$ & $33$\\
\hline
$70$ & $30$ & $33$ & $32$\\
\hline
$70$ & $31$ & $32$ & $31$\\
\hline
$70$ & $32$ & $31$ & $30$\\
\hline
$70$ & $33$ & $30$ & $29$\\
\hline
$70$ & $34$ & $29$ & $28$\\
\hline
$70$ & $35$ & $28$ & $27$\\
\hline
$70$ & $36$ & $27$ & $26$\\
\hline
$70$ & $37$ & $26$ & $25$\\
\hline
$70$ & $38$ & $25$ & $24$\\
\hline
$70$ & $39$ & $24$ & $23$\\
\hline
$70$ & $40$ & $23$ & $22$\\
\hline
$70$ & $41$ & $22$ & $21$\\
\hline
$70$ & $42$ & $21$ & $20$\\
\hline
$70$ & $43$ & $20$ & $19$\\
\hline
$70$ & $44$ & $19$ & $18$\\
\hline
$70$ & $45$ & $18$ & $17$\\
\hline
$70$ & $46$ & $17$ & $16$\\
\hline
$70$ & $47$ & $16$ & $15$\\
\hline
$70$ & $48$ & $15$ & $14$\\
\hline
$70$ & $49$ & $14$ & $13$\\
\hline
$70$ & $50$ & $13$ & $12$\\
\hline
\end{tabular}
\end{table}

Table \ref{f4} gives the new codes obtained from $\mathcal{X}_2$. The
codes have length $N$, dimension $K$ and minimum distance $D$. $D_m$ is
the lower bound on the minimum distance of codes from \cite{mint} with the same length and
dimension as the constructed generalised AG codes.
\section{Conclusion}

We have presented $41$ new codes  and $400$
improvements on constructible codes in $\mathbb{F}_{16}$  over the codes in
\cite{mint}. The construction method yields many good codes as shown
here,  as long as curves with many places of small degree are used,
however traditional search for good codes has focused primarily on
finding rational curves with many points. In order to obtain more codes
from generalised AG codes, curves with small genera and many places of small degree need to be found. 

\newpage

\bibliographystyle{IEEEtran}

\bibliography{IEEEabrv,references}

\end{document}